\documentclass[5p]{elsarticle}
\biboptions{sort&compress}
\usepackage{amsmath}
\usepackage{graphicx}
\usepackage{amssymb}
\usepackage{amsmath}

\newcommand{\NOT}{\textsc{not}}
\newcommand{\Fn}[1]{\ensuremath{\textrm{F}_{#1}}}
\newcommand{\Gn}[1]{\ensuremath{\textrm{G}_{#1}}}
\newcommand{\Pn}[1]{\ensuremath{\textrm{P}_{#1}}}
\newcommand{\Nn}[1]{\ensuremath{\textrm{N}_{#1}}}

\newcommand{\Sn}[1]{\ensuremath{\textrm{S}_{#1}}}
\newcommand{\Wn}[1]{\ensuremath{\textrm{W}_{#1}}}
\newcommand{\nuc}[2]{\ensuremath{{}^{#1}\textrm{#2}}}
\newcommand{\Bone}{\ensuremath{\textrm{B}_{1}}}

\begin{document}
\title{Further analysis of some symmetric and antisymmetric composite pulses for tackling pulse strength errors}
\author[oxford]{Sami Husain}
\author[okayama]{Minaru Kawamura}
\author[oxford]{Jonathan A. Jones\corref{cor1}}
\ead{jonathan.jones@qubit.org}
\address[oxford]{Centre for Quantum Computation, Clarendon Laboratory, University of Oxford, Parks Road, OX1 3PU, United Kingdom}
\address[okayama]{Electrical and Electronic Engineering, Okayama University of Science, 1-1 Ridai-cho, Okayama, 700-0005, Japan}
\cortext[cor1]{Corresponding author}
\date{\today}
\begin{abstract}
Composite pulses have found widespread use in both conventional Nuclear Magnetic Resonance experiments and in experimental quantum information processing to reduce the effects of systematic errors.  Here we describe several families of time symmetric and antisymmetric fully compensating composite pulses, inspired by the previous \Fn{n}, \Gn{n} and BB1 families family developed by Wimperis.  We describe families of composite $180^\circ$ pulses (\NOT\ gates) which exhibit unprecedented tolerance of pulse strength errors without unreasonable sensitivity to off-resonance errors, and related families with more exotic tailored responses.  Next we address the problem of extending these methods to other rotation angles, and discuss numerical results for $90^\circ$ pulses.  Finally we demonstrate the performance of some $90^\circ$ and $180^\circ$ pulses in NMR experiments.
\end{abstract}
\maketitle
\section{Introduction.}
NMR experiments suffer from a range of errors which can be traced back to imperfections in the radiofrequency control fields used to manipulate spin systems.  In particular the \Bone\ field is not infinitely strong, leading to off-resonance errors, and is not uniform across the sample, leading to pulse strength errors.  One particularly successful approach for tackling such errors is the use of composite pulses \cite{Levitt1979,Levitt1986}, in which a single rotation is replaced by a series of rotations chosen such that in the absence of errors the combined propagator implements the desired rotation, while in the presence of small errors the effects of the errors in individual rotations largely cancel one another.  Here we principally consider the problem of tackling pulse strength errors without introducing additional sensitivity to off-resonance errors.

Corresponding errors arise in experimental implementations of quantum information processing (QIP) \cite{Bennett2000,Jones2001a} where they are known as systematic errors \cite{Cummins2000}, to distinguish them from random errors arising from decoherence, and where there is considerable interest in performing extremely accurate unitary transformations on quantum systems in the presence of realistic experimental errors.  Again composite pulses provide a potential solution, but there are many significant differences between the design of composite pulses for quantum computing and conventional NMR.

Firstly, many traditional composite pulses, are designed to act correctly only on particular initial states, and so are not suitable for quantum computing, where fully compensating (Class A) composite pulses \cite{Levitt1986} have to be used \cite{Cummins2000,Cummins2003,McHugh2005,Xiao2006,Jones2011,Ichikawa2011,Ichikawa2012,Bando2013,Merrill2012}.  (While fully compensating pulses can be used for conventional NMR experiments, where they have the advantage that they can be immediately substituted for any simple pulse without the need for detailed analysis \cite{Levitt1986}, their use is normally excessive and better results can normally be obtained using pulses tailored to the relevant problem.)  Similarly it is necessary to use genuinely fully compensating pulses, rather than using phase cycling or gradient coherence pathway selection to suppress imperfections \cite{Odedra2012}.

Secondly, although composite pulses to suppress off-resonance errors, originally developed by Tycko \cite{Tycko1985}, have been used for QIP in NMR \cite{Cummins2000,Cummins2003} and SQUID \cite{Collin2004} experiments, tackling off-resonance errors is (with the exception of dynamic decoupling, discussed below) rarely of much interest in quantum computing, as the aim is to control a previously well characterised spin system, rather than to investigate an unknown system.  It is, however, generally desirable to ensure that the sensitivity to off-resonance errors is not greatly increased, and in particular is not significantly worse than that of a simple pulse.

Thirdly, it is useful to distinguish between composite pulses designed for single use, and pulses designed for use in a pulse train, such as decoupling sequences, where it may be desirable to combine relatively simple composite pulses with phase cycles and supercycles \cite{Levitt1981,Shaka1987a}.  In quantum computing most work has concentrated on single pulses, but in the field of dynamic decoupling \cite{Viola1999,Uhrig2007}, which is likely to play an important role in the design of quantum memories, the use of techniques adapted from conventional decoupling sequences has proved fruitful \cite{Souza2011b,Souza2012}.  Here we will only consider the case of single pulses.

Finally, quantum computing generally requires very accurate control in the presence of small or moderate errors, rather than moderately accurate control with very large errors.  (These two effects sometimes go together, but sometimes do not).  For this reason it is common to express the fidelity of the pulse as a Taylor series as described below, and to seek to suppress as many error terms as possible.  In particular there has been significant interest in the possibility of developing families of arbitrarily accurate composite pulses \cite{Brown2004,Brown2005}, although actually designing pulses by such methods is normally a complex problem \cite{Alway2007}.

Here we will show that it is straightforward to find robust $180^\circ$ pulses with arbitrary tolerance of pulse strength errors, but extending these ideas to pulses with other rotation angles is more challenging and only partially solved.  As with other composite pulses developed in NMR these can be extended to apply to spin--spin couplings \cite{Jones2003b} and can also be applied in a wide range of other quantum computing experiments \cite{Gulde2003,Morton2005a,Clayden2012,Ivanov2012}.  We will restrict ourselves to $\theta_0$ rotations about the $x$-axis of the Bloch sphere; rotations about other axes can be trivially derived from these by offsetting all pulse phases by the desired phase angle.

\section{Pulse fidelity}
It is convenient to characterize simple and composite pulses by their propagator fidelity \cite{Jones2011}
\begin{equation}
\mathcal{F}=\left|\textrm{tr}(VU^\dag)\right|/2
\label{eq:fiddef}
\end{equation}
where $V$ is the propagator of the pulse in the presence of errors and $U$ is the propagator for the ideal pulse (taking the absolute value is necessary in general to neglect the effects of global phases, and dividing by two simply normalises the fidelity).  Equivalently pulses can be categorized by their \textit{infidelity}, defined by $\mathcal{I}=1-\mathcal{F}$.

Much work to date has concentrated on pulse strength errors, which arise from errors in the strength of the \Bone\ field.  These are closely related to pulse length errors, and this name is often used instead, but a careful distinction between the two errors can and should be made \cite{Levitt1982}.  In particular the two cases are not quite the same in the presence of simultaneous off-resonance effects, as it is important to distinguish between measuring the off-resonance field as a fraction of the nominal \Bone\ field and as a fraction of the actual \Bone\ field \cite{Shaka1983}.  In reality \textit{both} types of error can be present: while naturally occurring errors usually arise from inhomogeneity in the \Bone\ field strength, many experimental demonstrations of composite pulses introduce additional deliberate errors by varying the pulse length \cite{Levitt1982}.

In the absence of off-resonance errors, pulse strength errors replace the rotation angle $\theta$ of each pulse by $\theta'=(1+\epsilon)\,\theta$, where $\epsilon$ is the fractional error.  The pulse fidelity can then be conveniently expanded as a Taylor series in $\epsilon$; for a simple pulse this leads to
\begin{equation}
\mathcal{F}=|\cos(\epsilon\theta/2)|=1-\epsilon^2\theta^2/8+O(\epsilon^4).
\end{equation}
The BB1 pulse sequence, originally developed by Wimperis \cite{Wimperis1994}, is effective in such cases; it replaces a $\theta_0$ pulse by the pulse sequence
\begin{equation}
\pi_\beta\,2\pi_{3\beta}\,\pi_\beta\,\theta_0\qquad\beta=\pm\arccos(-\theta/4\pi)
\label{eq:BB1}
\end{equation}
written with time running from left to right.  (The choice of sign is arbitrary but must be made consistently.)  The initial three pulse sequence performs no overall rotation in the absence of errors, but in the presence of pulse length errors generates a pure error term which largely cancels the error in the final $\theta_0$ pulse, suppressing both second and fourth order infidelities.  While Wimperis originally applied the correction sequence at the start of the composite pulse, as shown above, it can instead be placed at the end of the sequence, or indeed in the center of the sequence by splitting the $\theta_0$ pulse into two equal halves \cite{Cummins2003}.  We will initially only consider the case of $180^\circ$ pulses; these are used in NMR QIP to implement \NOT\ gates and in more conventional NMR to implement spin echoes and their generalisation dynamic decoupling \cite{Viola1999,Uhrig2007,Souza2011b,Souza2012}.  The fidelity of a BB1 $180^\circ$ pulse is
\begin{equation}
\mathcal{F}_\textit{BB1}=1-\epsilon^6\times5\,\pi^6/1024+O(\epsilon^8).
\end{equation}
Note that as usual a propagator infidelity of order $2n$ corresponds to an error term in the underlying propagator of order $n$ \cite{Jones2011}, and so the sixth-order sequence BB1 removes the first and second order errors from the propagator.

BB1 has found widespread use in quantum computing experiments, but there is interest in designing Class A pulses with even greater error suppression.  (While the approaches of arbitrary precision composite pulses have been used to design candidate pulses \cite{Brown2004,Brown2005,Alway2007}, those explored to date have not proved useful as their theoretically superior performance is not confirmed in experiments \cite{Xiao2006}.)  The most naive approach, constructing a higher order composite pulse by iteratively replacing each component pulse with a composite pulse, is rarely successful.  This is not surprising, as in general there is no reason to expect that the residual errors in a composite pulse should have the same form as the errors in a simple pulse, and so it is unlikely that the very same approach will work at different orders of iteration.

Instead methods for generating higher order pulses by recursive or iterative procedures (e.g., \cite{Shaka1983,Levitt1983,Tycko1984}) normally work directly on the error propagator \cite{Odedra2012b}.  An unusual partial exception is the \Sn{n} family of composite inversion pulses \cite{Tycko1984}; although this family was designed by considering the error propagator, the resulting sequences are in fact simply naive iterative expansions.  However the \Sn{n} family are not good Class A pulses, and so are not generally suitable for quantum computing.  More generally Tycko \textit{et al.} have considered the use of naive iterative expansions to design inversion and excitation pulses \cite{Tycko1985a}, and have discussed the conditions under which this approach is successful.

As we will discuss below, various families of Class A composite pulses described by Wimperis, can be described by a simple iterative process, and thus provide a simple route to arbitrary precision $180^\circ$ degree pulses for use in quantum computing.  Such pulses are also moderately robust at very large pulse strength errors, and so could in principle find applications in experiments using very inhomogeneous \Bone\ fields, such as phase-modulated rotating-frame imaging \cite{Allis1991} or single-sided NMR \cite{Blumich2008}, as well as effectively removing the need to calibrate RF pulse widths in walkup NMR experiments.  However in conventional NMR experiments it is likely that tailored approaches will be more effective.  While many suitable pulses have been developed using conventional approaches, we also draw attention to the robust inversion sequences developed by Torosov \cite{Torosov2011}, while in walkup NMR shaped pulses developed using optimal control theory \cite{Khaneja2005}, such as the Fanta4 pulses \cite{Nimbalkar2013} may also be of interest.

\section{Antisymmetric composite pulses}
The Wimperis error correcting sequence is time symmetric, and the BB1 sequence as a whole will similarly be symmetric if the correction sequence is placed in the middle of the main pulse. As we will see later such symmetric pulses have certain advantages, especially in the presence of off-resonance errors, and many composite pulses have this property.  Here we instead consider time antisymmetric composite pulses $180^\circ$ pulses, which can have particular advantages in some conventional NMR experiments \cite{Wimperis1991,Odedra2012}.  As we shall see, it is possible to combine many of the best features of the two approaches, by first designing antisymmetric composite pulses and subsequently converting these into corresponding symmetric pulses, or \textit{vice-versa}.

The design of antisymmetric composite $\pi$ pulses rests on the observation that, in the absence of errors, any time antisymmetric sequence of $\pi$ pulses such as
\begin{equation}
\pi_{-\phi_n}\dots\pi_{-\phi_1}\,\pi_0\,\pi_{\phi_1}\dots\pi_{\phi_n}
\label{eq:antisym}
\end{equation}
is equivalent to a simple $\pi_0$ pulse \cite{Wimperis1991}.  Thus the phases of the outer pulses, $\{\phi_1,\,\phi_2,\dots\phi_n\}$, can be adjusted at will, with the possibility of creating error tolerance. For the case of a five pulse sequence, with two controllable phase angles, the system can be solved analytically, and the second order infidelity term can be removed by choosing $\phi_1=\pm\phi$ and $\phi_2=\pm3\phi$ with $\phi=\arccos(-1/4)\approx104.5^\circ$. Once again the choice of sign is arbitrary, but must be made consistently, and in this case negating the phases corresponds to replacing the composite pulse with its time reversed form.  In passing we note that such antisymmetric pulses have the useful property that the rotation axis of the overall propagator always lies in the $xz$ plane \cite{Tycko1985a}, a fact which has considerable importance in some conventional NMR applications \cite{Odedra2012}, but which is not directly relevant in QIP.

Intriguingly these phases are identical to the BB1 phase angles for a $\pi$ pulse, and just like BB1 this pulse also suppresses the fourth order infidelity; indeed the fidelity of the antisymmetric pulse is \textit{identical} to that of the BB1 $\pi$ pulse.  This is not coincidental, but rather can be traced to two simple facts.  Firstly the fidelity definition, Eq.~\ref{eq:fiddef}, is invariant under cyclic reordering, and secondly we can use the identity
\begin{equation}
\pi_0\,\theta_\phi\equiv\theta_{-\phi}\pi_0
\label{eq:piswap}
\end{equation}
for a $\pi_0$ pulse and any other pulse.  Thus
\begin{align}
\mathcal{F}&=\textrm{tr}\left(\pi'_{-\phi_2}\,\pi'_{-\phi_1}\,\pi'_0\,\pi'_{\phi_1}\,\pi'_{\phi_2}\,\pi_0\right)/2\\
&=\textrm{tr}\left(\pi'_0\,\pi'_{\phi_1}\,\pi'_{\phi_2}\,\pi_0\,\pi'_{-\phi_2}\,\pi'_{-\phi_1}\right)/2\\
&=\textrm{tr}\left(\pi'_0\,\pi'_{\phi_1}\,\pi'_{\phi_2}\,\pi'_{\phi_2}\,\pi'_{\phi_1}\,\pi_0\right)/2
\end{align}
where $\pi'=\pi(1+\epsilon)$ as before.  Any antisymmetric sequence of five $\pi$ pulses can be converted into the same form as a BB1 pulse, and so the optimal error correcting sequences will occur at the same phase angles.  The ambiguity in the sign of the arccos function in Eq.~\ref{eq:BB1} is now seen to correspond to the equivalence of an antisymmetric pulse sequence and its time-reversed form.

This antisymmetric composite pulse was previously described by Wimperis \cite{Wimperis1991,Odedra2012}, who called it an \Fn1 pulse.  It was originally derived by using a toggling frame description of an antisymmetric composite pulse to isolate the error term, and then removing this term to first order \cite{Wimperis1991}.  \Fn1 is, however, simply the first member of the \Fn{n} family of pulses with rapidly growing error tolerance.  Wimperis explicitly described \Fn2, the second member of the family, and gave an outline prescription for designing further members.

Wimperis designed the family using an iterative expansion of the error propagator, but it is instead possible to proceed by a naive iterative expansion, effectively replacing each $\pi$ pulse in an \Fn1 sequence by an \Fn1 sequence.  It is, however, necessary to replace alternate pulses by ``standard'' \Fn1 sequences and their time-reversed equivalents.  This naive iterative approach to the design of Class A composite pulses is not usually successful, but its application to time antisymmetric sequences of $\pi$ pulses, and the need to alternate standard and time-reversed variants, can be understood by considering the composite pulse in the toggling frame \cite{Wimperis1991}.

By contrast with other approaches \cite{Brown2004,Brown2005,Alway2007} the \Fn{n} family provides a simple and direct method for explicitly constructing Class A composite pulses with arbitrary precision: the whole family can be easily described by listing the phases of successive $\pi$ pulses, with the iterative form
\begin{equation}
\phi_{n+1}=\{-3\phi+\phi_n,\,-\phi-\phi_n,\,\phi_n,\,\phi-\phi_n,\,3\phi+\phi_n\}
\label{eq:phin}
\end{equation}
with $\phi_0=0$ and $\phi=\arccos(-1/4)$.  Thus
\begin{equation}
\phi_1=\{-3,\,-1,\,0,\,1,\,3\}\times\phi
\end{equation}
reproducing \Fn1, and
\begin{multline}
\phi_2=\{-6,\,-4,\,-3,\,-2,\,0,\,2,\,0,\,-1,\,-2,\,-4,\,-3,\\
-1,\,\,0,\,1,\,3,\,4,\,2,\,1,\,0,\,-2,\,0,\,2,\,3,\,4,\,6\}\times\phi
\end{multline}
describes the 25 pulses making up \Fn2. In this notation the direct implementation of a \NOT\ gate as a single pulse can be considered as the family member \Fn0.  Note that the alternating pattern of plus and minus signs in Eq.~\ref{eq:phin} corresponds to the alternate use of standard and negative phase (time-reversed) pulses; this is necessary because the phase of the error term in the toggling frame is alternately positive and negative \cite{Wimperis1991}, and the phase steps of the component pulses must be alternately negated to follow this pattern.

The \Fn2 composite pulse suppresses all infidelity terms below eighteenth order, with
\begin{equation}
\mathcal{F}_{\Fn2}=1-\epsilon^{18}\times625\,\pi^{18}/2147483648+O(\epsilon^{20}),
\end{equation}
and the performance continues to improve rapidly for higher members of the family, as listed in Table~\ref{tab:F}.  (These results were obtained by direct calculation with pulse sequences of the form of Eq.~\ref{eq:phin}.) In general \Fn{n} is a sequence of $5^n$ pulses with a fidelity given empirically by
\begin{equation}
\mathcal{F}_{\Fn{n}}=1-\epsilon^{2q}\times5^{(q-1)/2}\times\pi^{2q}\times2^{(1-7q)/2}+O(\epsilon^{2q+2})
\label{eq:FFn}
\end{equation}
with $q=3^n$ (this expression has not been checked beyond \Fn5, which suppresses all infidelity terms below $486^\textit{th}$ order).  For the higher members of the family the first non-zero coefficient is very large, but has little effect at moderate $\epsilon$ as it is combined with an \textit{extremely} high power of $\epsilon$. Note that when evaluating the properties of sequences beyond \Fn2 it is essential to use either fully analytic methods or very high precision numerical calculations in order to capture fully the complex pattern of cancelations that leads to very high order error tolerances.

\begin{table}
\begin{tabular}{lllll}
\hline
$n$&pulses&order&coefficient&numerical value\\
\hline
0&1&$\epsilon^2$&$\pi^2/2^3$&$1.234$\\
1&5&$\epsilon^6$&$5\times\pi^6/2^{10}$&$4.694$\\
2&25&$\epsilon^{18}$&$5^5\times\pi^{18}/2^{31}$&$258.6$\\
3&125&$\epsilon^{54}$&$5^{13}\times\pi^{54}/2^{94}$&$4.324\times10^7$\\
4&625&$\epsilon^{162}$&$5^{40}\times\pi^{162}/2^{283}$&$2.021\times10^{23}$\\
\hline
\end{tabular}
\caption{Summary of the performance of \Fn{n} pulses for moderate values of $n$.  \Fn{n} pulses are only perfect at $\epsilon=0$ but near this point show infidelities of the orders shown.}
\label{tab:F}
\end{table}
\begin{figure}
\includegraphics{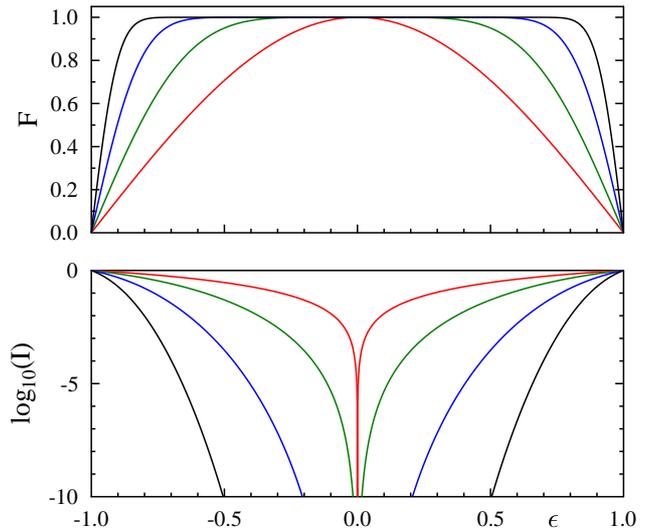}
\caption{Fidelity $\mathcal{F}$ and infidelity $\mathcal{I}$ as a function of pulse strength error $\epsilon$ for the \Fn{n} family of pulses from \Fn0 (plotted in red) to \Fn3 (plotted in black).}\label{fig:Fplot}
\end{figure}

The performance of some of the lower members of the \Fn{n} family is shown in Fig.~\ref{fig:Fplot}, clearly showing the very broad error tolerance achievable.  Theoretically these pulses allow astonishingly precise rotations to be performed, but in practice such extremely precise rotations are unlikely to be either required or achievable, and for this reason the infidelity plots shown here concentrate on infidelities around $10^{-5}$.  These practical limits arise because no experimental implementation will ever be completely perfect in every regard.  In particular, it is implausible that the required phase shifts can be produced precisely, and it is likely that relaxation during pulses will become a problem with the very long pulses required for high members of the family.

A similar approach can be used to describe the \Gn{n} family of composite pulses \cite{Wimperis1991}.  Rather than seeking to create a composite pulse with high fidelity over a central region, these pulses are designed to give perfect fidelity at certain particular error values, in the hope that this will lead to moderate error tolerance over a wide range of intermediate values.  (Error tolerance of this kind is not likely to be particularly useful for quantum information processing, but may find applications in more conventional NMR.)

The first non-trivial member, \Gn1, is obtained by forming an antisymmetric sequence of $\pi$ pulses as before and then choosing the phases to maximise the pulse fidelity at $\epsilon=\pm0.5$.  As before the problem can be solved analytically, giving the sequences of phases
\begin{equation}
\gamma_1=\{1,\,-2,\,0,\,2,\,-1\}\times\gamma
\end{equation}
with $\gamma=\pi/4$, a sequence which Wimperis called \Gn1.  The antisymmetric structure of the pulse guarantees that it will also be perfect at $\epsilon=0$, and the infidelity is quadratic in $\epsilon$ around these fixed points, with a shallow minimum in the fidelity around $\epsilon\approx0.28$.  Converting this sequence into a time symmetric form gives the Wimperis composite pulse BB2 \cite{Wimperis1994}.

\begin{figure}
\includegraphics{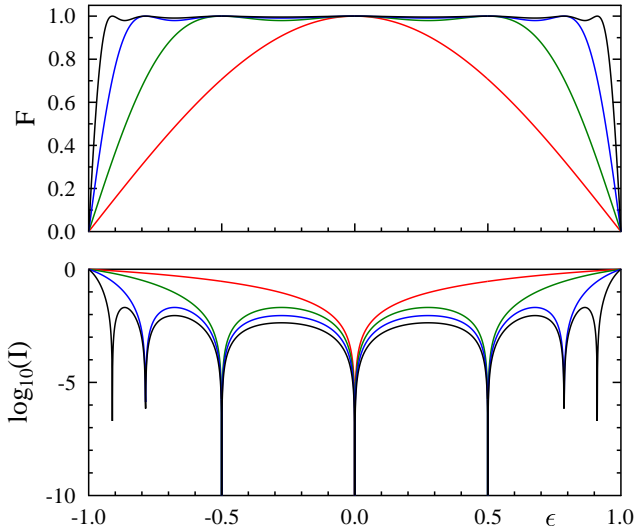}
\caption{Fidelity $\mathcal{F}$ and infidelity $\mathcal{I}$ as a function of pulse strength error $\epsilon$ for the \Gn{n} family of pulses from \Gn0 (red) to \Gn3 (black).  Note that only the middle three infidelity minima occur at rational values of $\epsilon$, and the outer minima are not well digitised.}\label{fig:Gplot}
\end{figure}

There is nothing special about the choice of $\epsilon=\pm0.5$ for the points of ideal fidelity, and close relatives of \Gn1 can be obtained for any value up to $0.8$.   However beyond $0.5$ the depth of the fidelity minimum increases rapidly, removing the desired broad error tolerance, while for small values the composite pulse sequence becomes very similar to \Fn1.  We will, therefore, stick to the original choice of $0.5$.  Once again this composite pulse can be expanded iteratively to get higher members of the \Gn{n} family.  The rule for pulse phases is entirely analogous to that for \Fn{n}, and is
\begin{equation}
\gamma_{n+1}=\{\gamma+\gamma_n,\,-2\gamma-\gamma_n,\,\gamma_n,\,2\gamma-\gamma_n,\,-\gamma+\gamma_n\}
\label{eq:gamn}
\end{equation}
with $\gamma_0=0$ and $\gamma=\pi/4$.  \Gn2 has perfect fidelity at $\epsilon=0$ and $\epsilon=0.5$, just like \Gn1, but also has perfect fidelity at two additional points at $\epsilon\approx\pm0.786$.  In general \Gn{n} has all the same perfect points as \Gn{n-1}, and a pair of additional points further out, as shown in Table~\ref{tab:G} and Fig.~\ref{fig:Gplot}.  Beyond \Gn1 there is no simple formula for the location of these points, which must be found numerically.
\begin{table}
\begin{tabular}{lllllll}
\hline
$n$&pulses&$\epsilon$\\
\hline
0&1&$0.000$\\
1&5&$0.000$&$\pm0.500$\\
2&25&$0.000$&$\pm0.500$&$\pm0.786$\\
3&125&$0.000$&$\pm0.500$&$\pm0.786$&$\pm0.911$\\
4&625&$0.000$&$\pm0.500$&$\pm0.786$&$\pm0.911$&$\pm0.963$\\
\hline
\end{tabular}
\caption{\Gn{n} pulses exhibit perfect fidelity at $2n+1$ points with the approximate values shown.}
\label{tab:G}
\end{table}

\section{Narrowband and passband pulses}
In addition to describing two symmetric families of broadband sequences, BB1 which corresponds to \Fn1 and BB2 which corresponds to \Gn1, Wimperis also described \cite{Wimperis1994} two families of narrowband sequences (which only give effective pulses when $\epsilon$ is close to zero) and two families of passband sequences (which act as error correcting composite pulses for moderate values of $\epsilon$, but act as moderately robust \textsc{identity} operators for values of $\epsilon$ near $\pm1$).  As usual we will consider the members corresponding to a nominal $\pi_0$ rotation.  These families are all time-symmetric, but as before we can convert them into time-antisymmetric sequences of $\pi$ pulses.  Having done this we can use the iterative approach to form higher members of the same family, and then (if desired) reverse the transformation to create time-symmetric sequences.

We will only discuss the narrowband family derived from NB1, which we call \Nn{n}, and the passband family derived from PB1, which we call \Pn{n}.  Applications of passband pulses in solid state NMR have recently been described \cite{Odedra2012a}, and we will demonstrate an application of narrowband pulses below.

\begin{figure}
\includegraphics{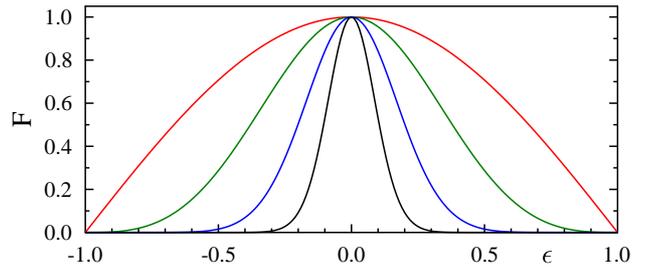}
\caption{Fidelity $\mathcal{F}$ as a function of pulse strength error $\epsilon$ for the \Nn{n} family of pulses from \Nn0 (plotted in red) to \Nn3 (plotted in black).}\label{fig:Nplot}
\end{figure}

An NB1 $\pi_0$ pulse takes the form
\begin{equation}
\pi_\nu\,2\pi_{-\nu}\,\pi_\nu\,\pi_0\qquad\nu=\arccos(-1/4)
\label{eq:NB1}
\end{equation}
and can be rearranged to get the antisymmetric pulse \Nn{1}, defined by the phases
\begin{equation}
\nu_1=\{1,\,-1,\,0,\,1,\,-1\}\times\nu.
\end{equation}
(A related pulse was previously discussed by Tycko \textit{et al.} \cite{Tycko1985a}, but they only considered its properties as an inversion pulse.)  As usual the higher members are defined by
\begin{equation}
\nu_{n+1}=\{\nu+\nu_n,\,-\nu-\nu_n,\,\nu_n,\,\nu-\nu_n,\,-\nu+\nu_n\}.
\label{eq:nun}
\end{equation}
All these pulses have second order infidelity around $\epsilon=0$, with the coefficient of this error term rising with increasing $n$
\begin{equation}
\mathcal{F}_{\Nn{n}}=1-\epsilon^{2}\times\pi^{2}/8\times(15/4)^n+O(\epsilon^{4}),
\end{equation}
creating a pulse which is only effective over increasingly narrow regions, as shown in Fig.~\ref{fig:Nplot}.

A PB1 $\pi_0$ pulse takes the form
\begin{equation}
2\pi_\psi\,4\pi_{-\psi}\,2\pi_\psi\,\pi_0\qquad\psi=\arccos(-1/8)
\label{eq:PB1}
\end{equation}
with a fidelity compared with an ideal $\pi$ pulse of
\begin{equation}
\mathcal{F}_{PB1}=1-\epsilon^6\times63\,\pi^6/1024+O(\epsilon^8)
\end{equation}
and a fidelity compared with an \textsc{identity} gate of
\begin{equation}
\mathcal{F}=1-(1+\epsilon)^4\times63\,\pi^4/512+O((1+\epsilon)^6)
\end{equation}
where the expansion of the fidelity has been taken around $\epsilon=-1$.  PB1 has slightly worse suppression of small errors than BB1, but gives very little excitation in regions of extreme pulse strength error.

\begin{figure}
\includegraphics{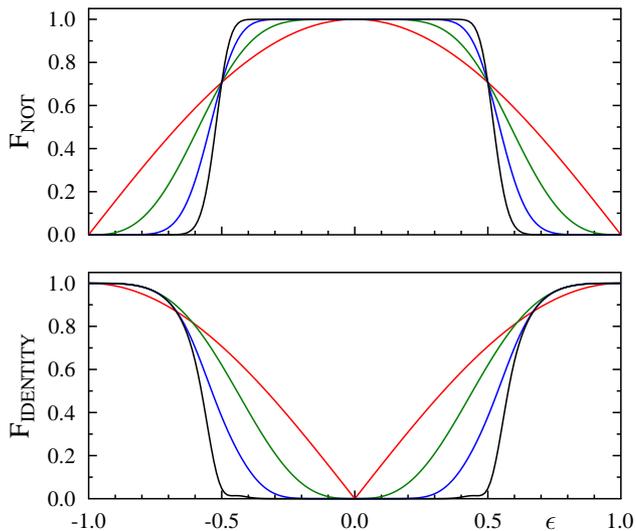}
\caption{Fidelity as a function of pulse strength error $\epsilon$ for the \Pn{n} family of composite pulses from \Pn0 (red) to \Pn3 (black).}\label{fig:Pplot}
\end{figure}

PB1 can be rearranged to give the \Pn1 sequence
\begin{equation}
\pi_{-\psi}\,\pi_{-\psi}\,\pi_{\psi}\,\pi_{\psi}\,\pi_0\,\pi_{-\psi}\,\pi_{-\psi}\,\pi_{\psi}\,\pi_{\psi}
\label{eq:P1}
\end{equation}
which has exactly the same fidelity but is now described as a time-antisymmetric sequence of 9 $\pi$ pulses.  (This pulse was previous discussed by Odedra and Wimperis \cite{Odedra2012a} under the name $\textrm{APB}_1$.)  This can be expanded iteratively to obtain \Pn2 and so on.  The \Pn2 sequence of 81 $\pi$ pulses has a fidelity
\begin{equation}
\mathcal{F}_{\Pn2}=1-\epsilon^{18}\times3^8\,7^4\,\pi^{18}/2^{31}+O(\epsilon^{20})
\end{equation}
with the expected eighteenth order behaviour. However the fidelity to an \textsc{identity} gate around $\epsilon=-1$ remains fourth order, with the same coefficient.  The behaviour of higher members of the family is shown in Fig.~\ref{fig:Pplot}, demonstrating increasingly robust passband behaviour.

\section{Combining pulses}
The results for F and G pulses discussed above are largely implicit in the earlier work of Wimperis \cite{Wimperis1991}, although he did not give an explicit form for \Fn{n} and \Gn{n}, or derive the remarkable error tolerances shown here, but the results for \Nn{n} are new, as are those for \Pn{n} for the sequences beyond \Pn1.  Our approach is not, however, confined to these four families.

\begin{figure}
\includegraphics{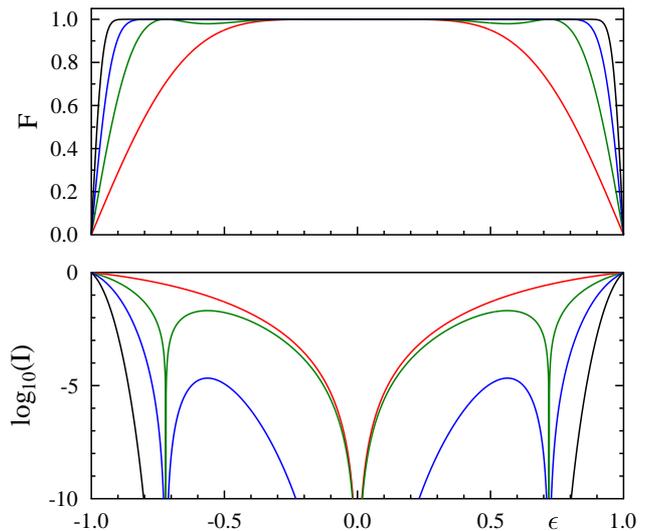}
\caption{Fidelity $\mathcal{F}$ and infidelity $\mathcal{I}$ as a function of pulse strength error $\epsilon$ for F (red), GF (green), FGF (blue) and FFGF (black) composite pulses.}\label{fig:FGFplot}
\end{figure}

All these composite pulses are built up iteratively, by replacing each $\pi$ pulse in a sequence by an error corrected $\pi$ pulse, but it is not necessary to use the same error-corrected pulse at each stage of the iteration, and the phase patterns described above can be applied more generally.  (Similar ideas have been explored in the design of composite inversion and excitation pulses \cite{Shaka1984,Tycko1985a}.)  In particular, consider the GF pulse obtained by applying the G pattern of phases to \Fn1
\begin{equation}
\psi_\textit{GF}=\{\gamma+\phi_1,\,-2\gamma-\phi_1,\,\phi_1,\,2\gamma-\phi_1,\,-\gamma+\phi_1\}
\end{equation}
to give a composite pulse made up from 25 $\pi$ pulses.  In common with \Fn1 this pulse has sixth order infidelity around $\epsilon=0$, and in common with \Gn1 this pulse has perfect fidelity at two additional values of $\epsilon$, although in this case they are moved out beyond 0.5, with quadratic error terms around these additional points.  This results in good error tolerance over the central region, and moderate error tolerance over a much wider region as shown in Fig.~\ref{fig:FGFplot}.  These new perfect points occur at $\epsilon\approx\pm0.720$ where the underlying \Fn1 pulse has a fidelity of $1/\sqrt{2}$, the same as the fidelity of a naive pulse at $\epsilon=0.5$.

In the same way the FG pulse can be obtained by applying the F pattern of phases to \Gn1
\begin{equation}
\psi_\textit{FG}=\{-3\phi+\gamma_1,\,-\phi-\gamma_1,\,\gamma_1,\,\phi-\gamma_1,\,3\phi+\gamma_1\}
\end{equation}
to get another 25 pulse sequence.  In common with \Gn1 this pulse exhibits perfect fidelity at $\epsilon=\pm0.5$, and in common with \Fn1 it has sixth order infidelity around $\epsilon=0$.  However this pulse is also sixth order around $\epsilon=\pm0.5$, giving good error tolerance over a wide region.

The process can be extended in the obvious way: for example the 125 pulse sequence FFG (or \Fn2G), obtained by applying the F phase pattern to FG, has perfect fidelity at $\epsilon=0$ and $\epsilon=\pm0.5$, and then shows eighteenth order infidelity around these three points.  Similarly the sequence FGF has eighteenth order infidelity around $\epsilon=0$ and sixth order infidelity around $\epsilon\approx0.720$ as shown in Fig.~\ref{fig:FGFplot}.  Any combination of F and G iterations can be applied: each application of G creates an additional pair of points at which the composite pulse has perfect fidelity, and each application of F triples the order of the infidelity around any existing perfect points.

\begin{figure}
\includegraphics{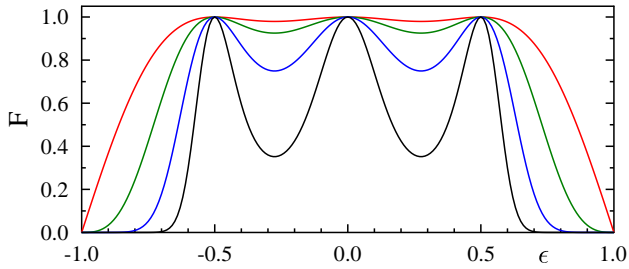}
\caption{Fidelity $\mathcal{F}$ as a function of pulse strength error $\epsilon$ for a G pulse (red), NG (green), \Nn{2}G (blue), and \Nn{3}G (black); the \Nn{3}G pulse is only effective near $\epsilon=0$ and $\epsilon=\pm0.5$.}\label{fig:NGplot}
\end{figure}

While the most obvious applications of this approach lie in the creation of broadband pulses, such as FGF, it is also possible to consider more exotic combinations.  For example, applying the N pattern of phases to a G pulse gives a family of composite pulse which in effect select for three particular control field strengths.  All such pulses show perfect fidelity at $\epsilon=0$ and $\epsilon=\pm0.5$, but as the N pattern of phases is repeatedly applied the fidelity away from these points is reduced, as shown in Fig.~\ref{fig:NGplot}.

\section{Off resonance errors}
This ability to sculpt the tolerance of pulse strength errors is remarkable, but it is important to check that it is not bought at the cost of increased sensitivity to other errors, in particular off-resonance errors.  Note that we do not consider here the design of pulses which are simultaneously resistant to off-resonance and pulse-length errors; discussions of this problem can be found elsewhere \cite{Alway2007,Ichikawa2011,Souza2011b,Souza2012,Odedra2012b,Bando2013}.

In the presence of simultaneous errors the propagator for a nominal $\theta_\phi$ pulse is
\begin{equation}
V=\exp(-\textrm{i}\theta[(1+\epsilon)(\sigma_x\cos\phi+\sigma_y\sin\phi)+f\sigma_z]/2)
\end{equation}
where $\epsilon$ is the pulse strength error as before, and $f$ is the \textit{off-resonance fraction} defined here as the ratio between the frequency offset of the transition from the pulse and the nominal rotation frequency of the pulse in the absence of any errors.  (Minor differences between the results shown here and those in previous work \cite{Cummins2003} reflect the fact that this earlier work considered pulse length errors rather than pulse strength errors.) In the absence of pulse strength errors the pulse fidelity is given by
\begin{equation}
\mathcal{F}=1-f^2\times\sin^2(\theta/2)/2+O(f^4)
\end{equation}
so the pulse infidelity is largest when $\theta=\pi$, for which $\mathcal{I}\approx f^2/2$, and smallest when $\theta=2n\pi$, where the first order error term completely cancels out.

In the absence of pulse length errors the BB1 composite pulse has the same sensitivity to small off-resonance errors as a single pulse; this favorable behavior can be traced to the fact that the time symmetric correction sequence is simply a $2\pi$ pulse inside another $2\pi$ pulse, and such nested sequences of $2\pi$ pulses have no first order error terms, so that the dominant source of infidelity is the main $\theta_0$ pulse.  This is not true for \Fn{n} and \Gn{n} pulses: direct calculations indicate that each iteration of the F phase sequence increases the size of the quadratic infidelity term by a factor of 16, while each iteration of G increases the infidelity by $9+2\sqrt{2}\approx11.8$.  Thus for an FG pulse, for example, the off-resonance infidelity is $\mathcal{I}\approx189.2f^2$.

This greatly increased sensitivity to off-resonance errors might appear a critical flaw, as it would render the higher members of the F and G families completely unusable in practice, but fortunately it can be easily overcome.  As noted above the BB1 composite $\pi$ pulse is very closely related to \Fn1, and can be obtained from it by moving the front part of the pulse sequence to the back with all the phases negated
\begin{equation}
\pi_{-3\phi}\,\pi_{-\phi}\,\pi_0\,\pi_\phi\,\pi_{3\phi}\longrightarrow
\pi_0\,\pi_\phi\,\pi_{3\phi}\,\pi_{3\phi}\,\pi_\phi.
\end{equation}
As implemented above all the pulses \textit{before} the main $\pi_0$ pulse have been moved, but it is also possible to move this pulse as well, or to move \textit{half} of the pulse to the back, resulting in a fully time-symmetric sequence, which has the advantage that the fidelity is then a purely even function of $f$.  This transformation can be applied to any time antisymmetric sequence of $\pi$ pulses, such as the F and G families, producing a nested sequence of $2\pi$ pulses: the extreme tolerance of pulse strength errors is retained, but now with little cost in sensitivity to off-resonance errors.

\begin{figure*}[t]
\includegraphics{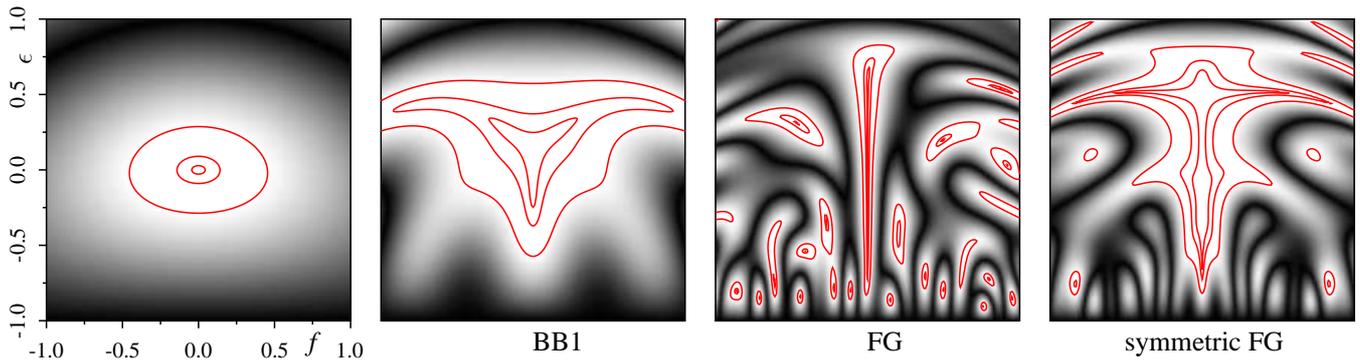}
\caption{Fidelity as a function of off resonance error $f$ and pulse strength error $\epsilon$ for \NOT\ gates implemented using simple pulses and BB1, FG, and symmetric BFG composite pulses; contours are drawn at fidelities of 0.9, 0.99 and 0.999.  Note that the subsidiary maxima in the FG and symmetric FG plots are not normally useful, and the key indicator is the size of the principal maximum in the centre of the plot.}\label{fig:BFGplot}
\end{figure*}

The calculations above only apply in the absence of pulse strength errors, and it is necessary to consider the effects of simultaneous errors which lead to cross terms in the fidelity expression.  These are best investigated numerically, as shown in Fig.~\ref{fig:BFGplot} which plots the fidelity of various simple and composite pulses over a wide range of errors.  While the simple FG sequence shows much greater tolerance of pulse strength errors than BB1, this is achieved at the cost of greatly increased sensitivity to off-resonance errors.  In contrast the time symmetric FG sequence achieves the same pulse strength tolerance as the antisymmetric FG sequence with only a modest increase in sensitivity to off-resonance errors compared with a simple pulse.

\section{Other pulse angles}
With the exception of the BB1 pulse, Eq.~\ref{eq:BB1}, all the results described above only apply to $\pi_0$ pulses (\NOT\ gates), and it would obviously be desirable to extend these to a wider range of pulse angles (the extension to $\pi$ pulses with other phase angles is trivially achieved by offsetting all the phases in a $\pi_0$ composite pulse by the desired phase angle).  In particular it would be helpful to extend these methods to $\pi/2$ pulses, both for applications in conventional NMR and in QIP. Note that in QIP the set of $\pi/2$ pulses with arbitrary phases is universal for single qubit gates, and in NMR systems a robust $\pi/2$ rotation can be extended by analogy to produce a controlled-\NOT\ gate \cite{Jones2003b}, and thus robust universal quantum logic.

This process might seem simple, as the BB1 composite pulse, Eq.~\ref{eq:BB1}, works for any rotation angle, and the BB1 $\pi_0$ pulse can be converted into an \Fn1 pulse. Such optimism is, however, unjustified, as the identity in Eq.~\ref{eq:piswap} only applies for $\pi$ pulses, and so the transformation can not be applied more generally.  Furthermore the key property of antisymmetric pulses of the form of Eq.~\ref{eq:antisym} only applies when the pulses are $\pi$ pulses.  For example, the sequence
\begin{equation}
\theta_{-\phi_2}\,\theta_{-\phi_1}\,\theta_0\,\theta_{\phi_1}\,\theta_{\phi_2}\qquad\theta\ne\pi
\label{eq:antisym90}
\end{equation}
is only equivalent to $\theta_0$ if $\phi_2=\phi_1+\pi$, and in general it is necessary to introduce similar relationships between the pulse phase angles.  While we have been able to develop some low order antisymmetric $\pi/2$ pulses, for applications in QIP these sequences are inferior to the time symmetric sequences discussed below, and are not considered further here.

We begin by rewriting the correction sequence at the start of the BB1 pulse, Eq.~\ref{eq:BB1}, in the general time-symmetric form
\begin{equation}
\pi_{\phi_1}\,\pi_{\phi_2}\,\pi_{\phi_2}\,\pi_{\phi_1}
\label{eq:W1}
\end{equation}
which we refer to as a \Wn1 sequence.  Clearly such a pulse can be summarised by listing the two phase angles, $\phi_1$ and $\phi_2$, which can be adjusted to optimise the fidelity of the combined pulse by removing terms in the Taylor series expansion.  In particular it is possible to eliminate the second and fourth order error terms, and for a \Wn1 $90^\circ$ pulse this occurs at $\phi_2=3\phi_1$ and $\phi_1=\arccos(-1/8)$, the well known BB1 result \cite{Wimperis1994}.  The process can then be extended by using larger numbers of adjustable phase angles and seeking to remove higher order terms.

With only two adjustable phases it is straightforward to tackle the problem analytically, seeking to set Taylor series coefficients to zero, but as the number of pulses is increased this method becomes intractable.  Instead we have adopted a numerical approach, effectively replacing the solution of simultaneous non-linear equations with a minimisation problem by seeking to minimise the sum of squares of the first $n$ coefficients: if this sum of squares can be reduced to a value indistinguishable from zero then the corresponding terms have been effectively eliminated from the series.  (This approach is known to be unreliable in general \cite{NumRec1992}, but seems to work fairly well in this particular case).  Similar ideas have been explored elsewhere \cite{Torosov2011,Ivanov2012}.

Our numerical searches show that adding a third adjustable phase does not help; in particular the sixth order error cannot be eliminated.  However adding a fourth adjustable phase permits \textit{both} the sixth order and the eight order terms to be eliminated with a sequence of eight $\pi$ pulses.  This pattern continues throughout the range of searches we have conducted: the \Wn{n} sequence of $4n$ pulses with $2n$ adjustable phases
\begin{equation}
\pi_{\phi_1}\,\pi_{\phi_2}\,\dots\,\pi_{\phi_{2n}}\,\pi_{\phi_{2n}}\,\dots\,\pi_{\phi_2}\,\pi_{\phi_1}
\label{eq:Wn}
\end{equation}
permits all the error terms up to order $4n$ to be removed.  For the case of \Wn1 there is only a single solution (the well known BB1 composite pulse), neglecting the trivial variant formed by negating all the pulse phases, but for \Wn2 there are two distinct solutions.  These appear in theory to be effectively identical; in particular the tenth order coefficient has the same size in the two cases.  At \Wn3 the search becomes computationally challenging, but once again two distinct solutions have been located.  All these solutions for $90^\circ$ pulses are listed in Table~\ref{tab:Wn90}.  Preliminary searches indicate that at least one \Wn4 sequence exists, but this has not been accurately located and it is not yet known whether any other sequences exist.
\begin{table}
\begin{tabular}{rrrrrrr}
\hline
$90^\circ$&$\phi_1$&$\phi_2$&$\phi_3$&$\phi_4$&$\phi_5$&$\phi_6$\\
\hline
\Wn1& 97.2&291.5&&&&\\
\Wn2& 84.3&162.0&345.5&286.7&&\\
\Wn2&132.3&339.1& 26.4&222.2&&\\
\Wn3& 22.0&186.0& 89.3&319.2&178.2&325.7\\
\Wn3& 79.9&119.2&257.3& 81.0&308.4&286.9\\
\hline
\end{tabular}
\caption{Composite $90^\circ$ pulses: the \Wn{n} correction sequence contains $4n$ pulses with $2n$ adjustable phases, and the phases (in degrees) listed permit all error terms up to order $4n$ to be eliminated when the correction sequence is combined with a naive $90^\circ$ pulse.}
\label{tab:Wn90}
\end{table}

This approach can of course be extended from $90^\circ$ pulses to other angles; in particular it is possible to find phases for $180^\circ$ pulses, as listed in Table~\ref{tab:Wn180}.
\begin{table}
\begin{tabular}{rrrrrrr}
\hline
$180^\circ$&$\phi_1$&$\phi_2$&$\phi_3$&$\phi_4$&$\phi_5$&$\phi_6$\\
\hline
\Wn1&104.5&313.4&&&&\\
\Wn2& 79.2&193.4& 24.9&307.5&&\\
\Wn2&130.0&  1.5& 56.7&259.9&&\\
\Wn3&341.9&147.9&100.5&355.9&207.3&339.4\\
\Wn3& 69.5&141.7&289.4&121.4&350.1&307.3\\
\hline
\end{tabular}
\caption{Composite $180^\circ$ pulses: the  phases (in degrees) listed permit all error terms up to order $4n$ to be eliminated when the \Wn{n} correction sequence is combined with a naive $180^\circ$ pulse.}
\label{tab:Wn180}
\end{table}
These composite $180^\circ$ pulses can then be converted into antisymmetric forms, and interestingly these appear to have the same property as \Fn{1} pulses, allowing them to be recursively nested.  However this point is not explored further here.

\section{Experiments}
Experiments were carried out in Oxford using a Varian Unity Inova system with a nominal \nuc{1}{H} frequency of 600\,MHz to study a sample of HOD doped with $\textrm{GdCl}_3$ to reduce the relaxation times. Experiments in Okayama used a homebuilt system with a \nuc{19}{F} frequency of 376\,MHz to study a sample of perfluorobenzene dissolved in deuterated benzene.  These two systems differ substantially in experimental detail, potentially allowing the effects of different experimental errors to be distinguished.  As our systems should not suffer from significant off-resonance errors it is possible to introduce pulse rotation angle errors either by adjusting the pulse length (pulse-length errors) or by adjusting the \Bone\ field strength (pulse strength errors) \cite{Levitt1982}.  We chose to introduce deliberate errors by adjusting the pulse length, but as discussed below the system also contains underlying pulse strength errors. 

\begin{figure}
\includegraphics{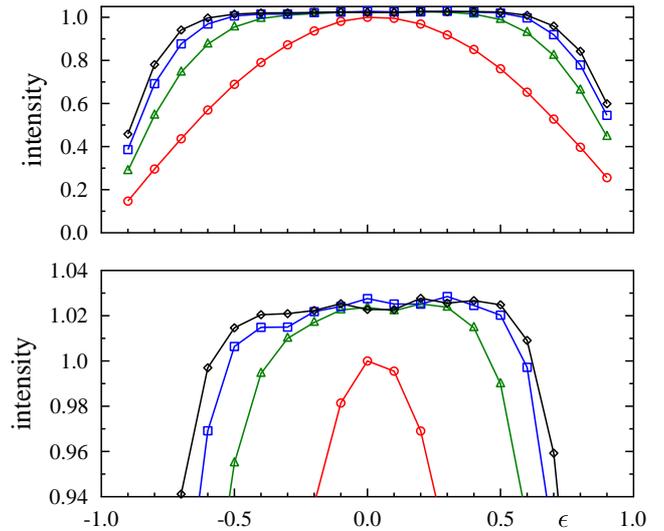}
\caption{Experimental signal intensity (arbitrary units) as a function of pulse length error $\epsilon$ for naive (red circles), \Wn1 (green triangles), \Wn2 (blue squares), and \Wn3 (black diamonds) $90^\circ$ pulses used as excitation pulses.  The lines simply join the experimental points and are plotted to guide the eye.  The lower panel shows a vertical expansion of the upper panel.}\label{fig:Wndata}
\end{figure}

We begin by demonstrating the use of composite $90^\circ$ pulses based on \Wn{n} correction sequences; for simplicity we simply show their performance as excitation pulses, but as Class A composite pulses the performance should be very similar for other initial states. The \Bone\ field strength on the Oxford system was carefully adjusted to 25\,kHz, corresponding to a nominal $90^\circ$ pulse length of $10\,\mu\textrm{s}$, and then naive and composite pulses were applied using pulse lengths varying between $1$ and $19\,\mu\textrm{s}$. The signal intensity was determined by integration, and as the vertical scale is largely arbitrary all intensities were normalised by dividing them by the intensity from a naive pulse with $\epsilon=0$.  The results are shown in Fig.~\ref{fig:Wndata}, with a vertical expansion in the lower panel; note that in this figure the lines simply join the experimental data points and are plotted to guide the eye.  All experiments used the first of the the two choices for the \Wn2 and \Wn3 sequences; initial experiments (data not shown) indicated that very similar results were obtained for the other two sequences.

It is clear that the \Wn{n} composite pulses perform largely as expected, with wider error tolerance for higher values of $n$.  However closer examination of the results, aided by the vertical expansion in the lower panel, shows two further effects.  Firstly the overall signal intensity is larger with composite pulses than with naive pulses, and secondly all four plots lean to the right, showing higher intensities for positive values of $\epsilon$ in comparison with the equivalent negative values.  Both of these effects can be ascribed to additional pulse strength errors arising from \Bone\ inhomogeneity.  The visible increase in overall intensity indicates that a significant fraction of the sample experiences a \Bone\ field substantially different from the central value, while the lean to the right indicates that these regions experience a \textit{smaller} \Bone\ field than the bulk of the sample.  This behaviour is expected for traditional RF coil designs.

\begin{figure}
\includegraphics{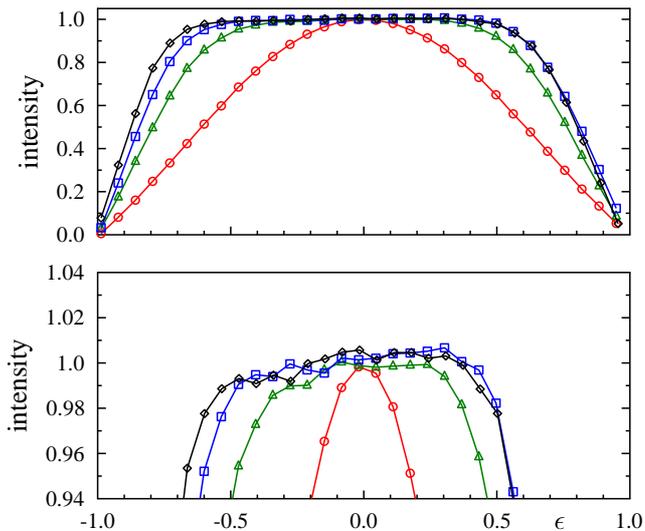}
\caption{Experimental signal intensity (arbitrary units) as a function of pulse length error $\epsilon$ for naive (red circles), \Wn1 (green triangles), \Wn2 (blue squares), and \Wn3 (black diamonds) $90^\circ$ pulses used as excitation pulses with a small spherical sample.
}\label{fig:WOkayamadata}
\end{figure}

Attempts were made to reduce these effects in experiments on the Oka\-yama system by replacing the conventional sample tube by a small spherical sample in the centre of the RF coil, which should reduce the RF inhomogeneity over the sample.  The results, shown in Fig.~\ref{fig:WOkayamadata}, indicate partial success. In particular the increase in overall intensity is reduced, but a lean to the right remains visible.

To make further progress we used narrowband $\pi$ pulses to select signal from regions of the sample with a highly homogeneous \Bone\ field strength. The double pulse field gradient spin echo (DPFGSE) sequence \cite{Hwang1995} is effective in selecting for spins which experience a $180^\circ$ pulse at the centre of each spin echo.  The classic use is to suppress water signals by combining a hard $\pi_x$ pulse with a soft $\pi_{-x}$ pulse which selects water transitions, but it can be used more widely to select subsets of spins.  Here we use an \Nn{2} composite $\pi$ pulse which only excites spins where the \Bone\ field strength is close to its nominal value, and compare the results with those from an \Fn{2} pulse, which should pass a very wide range of field strengths.  (This is simpler than comparing the \Nn{2} filtered data with spectra acquired without a selective filter, as the filtration process also causes signal loss due to the effects of diffusion \cite{Stejskal1965}.)

\begin{figure}
\includegraphics{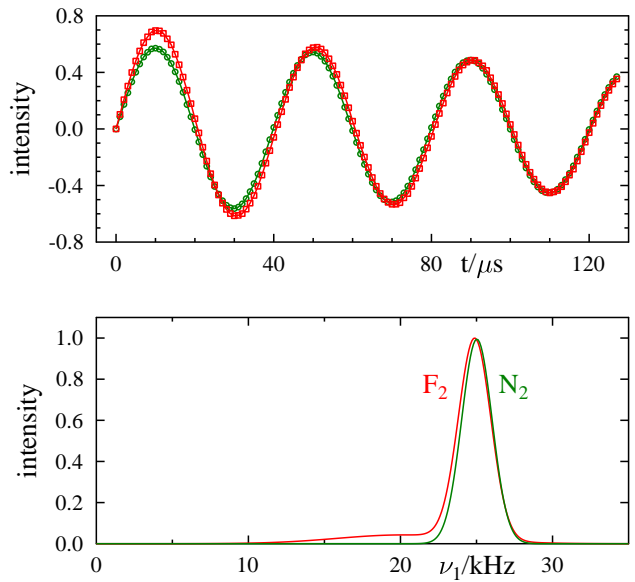}
\caption{Nutation experiments to measure RF inhomogeneity.  In the upper panel the experimental signal intensity (arbitrary units) is plotted as a function of pulse length with \Fn{2} (red) and \Nn{2} (green) DPFGSE filtration sequences.  The points show experimental values while lines show fitted curves.  The \Nn2 data is well described by a single narrow Gaussian component centred at 25\,kHz, while the \Fn2 data requires a two Gaussian fit, revealing a broad component centred around 20\,kHz.  The lower panel shows the distribution of nutation frequencies corresponding to these fits.}\label{fig:pwFN}
\end{figure}

These filters were explored using a nutation sequence to measure RF inhomogeneity, as shown in Fig.~\ref{fig:pwFN}.  The nominal \Bone\ nutation frequency was adjusted to 25\,kHz by adjusting the RF power until a 40\,$\mu$s pulse corresponded to a notation of $360^\circ$, and the nutation time varied between 0 and 127\,$\mu$s.  The \Nn2 data is well described by a single narrow Gaussian distribution of nutation frequencies, with a central value of 25\,kHz and a width of 2.5\,kHz, but the \Fn2 data needs a second Gaussian component, centred around 20\,kHz with a width of about 10\,kHz.  This low frequency broad component leads to the deviations from ideal behaviour seen in Fig.~\ref{fig:Wndata} and discussed above.

\begin{figure}
\includegraphics{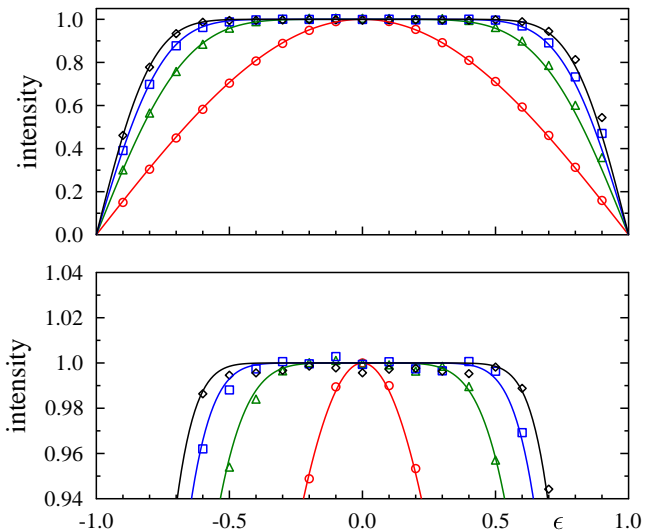}
\caption{Experimental signal intensity (arbitrary units) as a function of pulse length error $\epsilon$ for naive (red circles), \Wn1 (green triangles), \Wn2 (blue squares), and \Wn3 (black diamonds) $90^\circ$ pulses used as excitation pulses followed by the use of a DPFGSE \Nn2 sequence to select a small region with high \Bone\ homogeneity.  The points show experimental values while lines show the theoretical curves.  The lower panel shows a vertical expansion of the upper panel.}\label{fig:WN2data}
\end{figure}

Finally we show the effect of combining a \Wn{n} excitation pulse with a \Nn{2} filtration sequence.  In this case the aggravating pulse strength errors arising from RF inhomogeneity should be well suppressed, leaving only the artificially imposed pulse length errors.  The result, shown in Fig.~\ref{fig:WN2data} confirms this, with the experimental data points lying very close to theoretical predictions.

Note that in this plot the theoretical curves are simply plotted directly on the same scale as the experimental data, rather than being optimised to fit this data, showing the very close match between theory and experiment.  The remaining mismatch is most severe for the \Wn3 experiment, and may partly reflect spin--spin relaxation during this long composite pulse.  Remaining errors are likely to be due to imperfections in the phase-control of the RF source, and slow drifts in the power of the RF amplifier.

\section{Conclusions}
The ability to interconvert time symmetric and time antisymmetric $\pi$ pulses allows us to interpret previous results with a common framework, and then to extend these results to produce new analytic families of composite $\pi$ pulses with unprecedented tolerance of pulse strength errors at little or no cost in sensitivity to off-resonance errors.  This approach is currently confined to $\pi$ pulses, and cannot be applied to pulses with other rotation angles.  It is, however, possible to use numeric methods to design symmetric composite pulses for arbitrary angles, and several families have been located.  Experimental implementations confirm that these pulses work very much as expected, but it is necessary to perform these experiments carefully in order to avoid confusion arising from multiple sources of pulse strength error.

\section*{Acknowledgments}
We thank the UK EPSRC and BBSRC for financial support.  We are grateful to Stephen Jones, Ben Rowland and Steve Wimperis for helpful conversations.  We thank two reviewers for useful suggestions.

\bibliographystyle{elsarticle-num}
\bibliography{../../../all}

\end{document}